\documentclass[12pt]{JHEP3}

\usepackage{amsmath,epsfig}
\usepackage{amssymb,amsfonts}
\usepackage{latexsym}
\usepackage[latin1]{inputenc}
\usepackage{subeqnarray}
\usepackage{xcolor}

\usepackage{graphicx}
\usepackage{longtable}

\relax

\def\be{\begin{equation}}
\def\ee{\end{equation}}
\def\bea{\begin{eqnarray}}
\def\eea{\end{eqnarray}}

\newcommand\fverb{\setbox\pippobox=\hbox\bgroup\verb}
\newcommand\fverbdo{\egroup\medskip\noindent%
                        \fbox{\unhbox\pippobox}\ }
\newcommand\fverbit{\egroup\item[\fbox{\unhbox\pippobox}]}

\newcommand{\bear}{\begin{eqnarray}}

\newcommand{\eear}{\end{eqnarray}}

\newcommand{\bsea}{\begin{subeqnarray}}
\newcommand{\esea}{\end{subeqnarray}}
\newbox\pippobox

\def\6{\partial}

\def\a{\alpha}
\def\g{\gamma}

\def\pa{\partial}

\def\s{\sigma}

\def\sq
\def\a{\alpha}

\def\hri#1#2{\href{http://arxiv.org/abs/#1}{[ArXiv:#1]#2}}

\title{\huge Hyperscaling Violating Solutions in Generalised EMD Theory}
\author{\Large  Li Li$^{a,b,c}$\\
~\\
~\\
$^a$\href{http://hep.physics.uoc.gr}{Crete Center for Theoretical Physics, Institute for Theoretical and Computational Physics},
Department of Physics, University of Crete, 71003 Heraklion, Greece.
~\\
~\\
$^b$Crete Center for Quantum Complexity and Nanotechnology,
Department of Physics, University of Crete, 71003 Heraklion, Greece.\\
~\\
$^c$ Department of Physics, Lehigh University, Bethlehem, PA, 18018, USA.
~\\
~\\
E-mail: \email{lil416@lehigh.edu}

}

\preprint{CCTP-2016-12\\
CCQCN-2016-160}

\abstract{This short note is devoted to deriving scaling but hyperscaling violating solutions in a generalised Einstein-Maxwell-Dilaton theory with an arbitrary number of scalars and vectors.  We obtain analytic solutions in some special case and discuss the physical constraints on the allowed parameter range in order to have a well-defined holographic ground-state solution. }


\begin{document}

\section{Introduction}
\label{intro}

The application of the holographic duality towards understanding theories that are at finite density and may be in the universality class of strongly coupled systems has been made much progress~\cite{Jan,Ammon:2015wua}. Of particular interests is quantum criticality, which is crucial for interpreting a wide variety of experiments. A large class of critical points in condensed matter are characterised by two scaling exponents, known as the dynamical critical exponent $z$ and the hyperscaling violation exponent $\theta$.

Such exponents appear in holographic saddle point solutions. Lifshitz scaling solutions have been discussed first in~\cite{Kachru:2008yh} (for a recent review see~\cite{Taylor:2015glc}), while hyperscaling violating solutions were recognized in~\cite{Charmousis:2010zz,Gouteraux:2011ce}.

The duality provides a natural framework to describe those quantum critical systems. The metric in the gravitational dual description takes the form
\begin{equation}
\label{hsvmetric}
ds^2=r^{\frac{2\theta}{d}}\left(-\frac{dt^2}{r^{2z}}+\frac{L^2 dr^2+d\vec x^2}{r^2}\right)\,,
\end{equation}
with $d\vec x^2=dx_1^2+\cdots+d x_{d}^2$ and $d$ the number of spatial dimensions in the field theory. The scaling geometry possesses the property
\begin{equation}
r\rightarrow \lambda\, r,\quad t\rightarrow \lambda^z \,t,\quad x_i\rightarrow \lambda\,x_i,\quad ds^2\rightarrow \lambda^{\frac{2\theta}{d}}\,ds^2,
\end{equation}
where $\lambda$ is a dimensionless constant. Therefore, $z$ characterises the deviation from the Lorentz invariant and $\theta$ characterises the deviation from the scale invariant limit.
Aspects of hyperscaling violating geometry and its realisation in various gravity models have been widely discussed in the literature, see for example~\cite{Charmousis:2010zz,Gouteraux:2011ce, Dong:2012se,Gath:2012pg,Bueno:2012vx,Chemissany:2014xsa,Ghodrati:2014spa} and references therein. Those geometries with hyperscaling violation are usually considered as the infrared (IR) limit of some kind of bulk solutions that asymptotically approach AdS in the boundary. Due to the presence of nontrivial scaling exponents, there are novel behaviours relative to the AdS counterpart, see for example~\cite{Hartnoll:2009ns,Lee:2010ii,Iizuka:2011hg,Ogawa:2011bz,Bueno:2014oua,Hartnoll:2012wm,Iizuka:2013ag,Lucas:2014zea,Karch:2014mba,Li:2014xia,Kiritsis:2015oxa,Kiritsis:2015yna,Amoretti:2016cad,Cremonini:2016bqw,Salvio:2013jia,Ge:2016sel}. In particular, it was found that the recent pnictide data~\cite{Hayers:2014} can be well described by using holographic DBI magnetoresistance at quantum criticality with hyperscaling violation~\cite{Kiritsis:2016cpm}.

It is well known that hyperscaling violating solutions can be generated in the Einstein-Maxwell-Dilaton (EMD) theory where gravity couples to one real neutral scalar and one U(1) gauge field~\cite{Charmousis:2010zz,Gouteraux:2011ce}. We would like to generalise the simple EMD theory to involve an arbitrary number of scalars and vectors.  We will consider a bottom-up theory where the theory parameters can be turned continuously at the level of effective holographic theory.
Our motivation are two folds. Firstly, such kind of theory is common by consistent truncation of various supergravity theories in higher dimensions~\cite{Kiritsis:2016cpm,Ferrara:1997tw,Cvetic:1999xp,Donos:2010tu,Gouteraux:2011qh,Perlmutter:2012he}. We would like to describe possible geometries in those general setups. Such theory would then be either embedded in string theory/supergravity, or asymptotic to AdS. On the other hand, there have been recently a number of holographic models using multiple vectors and scalars, as they typically lead to richer physics~\cite{Donos:2012js,Kiritsis:2015hoa,Seo:2016vks,Cremonini:2016rbd}. For example, multiple U(1) gauge fields in the bulk will source multiple conserved currents in the dual field theory. Conductivities in such case have been discussed in~\cite{Sun:2013wpa,Cremonini:2016avj}, although their physical interpretations are not yet very clear.

We discuss the conditions for the existence of purely scaling geometry in our theory. It turns out that the existence of scaling solutions in general  imposes non-trivial constraints on theory parameters. Then we use the established formulae to the special case where the scalars take the standard kinetic term. We can find exact black brane solutions with arbitrary values of hyperscaling violation exponent $\theta$ and dynamical exponent $z$.

The rest of the paper is organised as follows. In section~\ref{setup} we introduce the gravity theory and derive the equations of motion. Section~\ref{solutions} is devoted to discussing the hyperscaling violating geometry. We give the conditions for the existence of such scaling solutions, which in general can not be solved analytically due to the scalar metric in front of the kinetic terms of scalars. Section~\ref{analytic} presents a set of exact solutions at extremal case as well as finite temperature case.  The constraints on the parameter range of $(\theta,z)$ are discussed in more details. An example from a top-down setup by using toroidal compactifications is given. We conclude in section~\ref{conclusion}.

\section{The General Theory and Equations of Motion}\label{setup}
We consider an effective gravitational theory that involves an arbitrary number of scalars and vector fields at the two-derivative level. The action reads
\begin{equation}\label{action}
S=\int d^{d+2}x\sqrt{-g}\left[\mathcal{R}-{1\over 2}\sum_{i,j=1}^M \mathcal{G}_{ij}(\phi)\nabla_\mu\phi_i \nabla^\mu\phi_j+V(\phi)-{1\over 4}\sum_{I=1}^N Z_I(\phi) F_I^2\right],
\end{equation}
which contains $M$ scalars $\phi_i$ and $N$ massless vectors $A_I$. We also generalize it a bit by allowing a non-trivial symmetric metric $\mathcal{G}_{ij}(\phi)$ for the scalars.

From the action~\eqref{action} we derive the equations of motion for the scalar $\phi_i$
\begin{equation}\label{eomphi}
\nabla_\mu\left(\sum_{j=1}^M\mathcal{G}_{ij}(\phi)\nabla^\mu \phi_j \right)-{1\over 2}\sum_{j,k=1}^M \frac{\partial\mathcal{G}_{jk}(\phi)}{\partial\phi_i} \nabla_\mu\phi_j \nabla^\mu\phi_k+\frac{\partial V(\phi)}{\partial\phi_i}-{1\over 4}\sum_{I=1}^N \frac{\partial Z_I(\phi)}{\partial\phi_i} F_I^2=0\,,
\end{equation}
and vector $A_I$
\begin{equation}\label{eomA}
\nabla_\mu(Z_I(\phi)F_I^{\mu\nu})=0\,,
\end{equation}
with $ i,j,k=1,\dots,M$ and $I=1,\dots,N$.
The equations of motion for the metric $g_{\mu\nu}$ are given by
\begin{equation}\label{eomg}
\begin{split}
\mathcal{R}_{\mu\nu}-{1\over 2}\mathcal{R}g_{\mu\nu}=&{1\over 2}\sum_{i,j=1}^M \mathcal{G}_{ij}(\phi)\left(\nabla_\mu\phi_i \nabla_\nu\phi_j-{1\over 2} g_{\mu\nu}\nabla_\rho\phi_i \nabla^\rho\phi_j\right)\\ +&{1\over 2}g_{\mu\nu}V(\phi)+{1\over 2}\sum_{I=1}^N Z_I(\phi) \left(F_{I\mu\rho}{F_{I\nu}}^{\rho}-{1\over 4}g_{\mu\nu}F_I^2\right)\,.
\end{split}
\end{equation}

We are interested in the hyperscaling violating solution in the generalised EMD theory. We further simplify the discussion by specialising to the diagonal scalar metric case, i.e., only turn on the diagonal metric $\mathcal{G}_{ii}(\phi)$. We approximate the scalar couplings have exponential asymptotics as in supergravity,
\begin{equation}
\mathcal{G}_{ii}\sim e^{\vec \tau_i\cdot \vec\phi},\quad V\sim V_0~e^{-\vec\delta\cdot \vec \phi},\quad Z_I\sim e^{\vec \gamma_I\cdot \vec \phi},
\end{equation}
with $V_0$ a positive constant. Here we have used a vector notation for $M$ scalars with $\vec\phi=(\phi_1,\phi_2,\cdots,\phi_M)$. So the theory we are considering depends on $(M+N+1)$ $M$-vectors $\vec \tau_i$, $\vec \gamma_I$ and $\vec\delta$. Those vectors will be related to the scaling exponents of the solutions, i.e.,  $z$ and $\theta$.

In this note we focus on the case with two spatial boundary dimensions ($d=2$) for simplicity, but our discussion can be generalised to higher dimensions straightforwardly. 
For the  homogeneous and isotropic case the bulk metric as well as matter part takes the generic form,
\begin{eqnarray}\label{fullansatz}
\begin{split}
ds^2=-D(r)dt^2+B(r)dr^2+C(r) (dx_1^2+dx_2^2)\,,\\
\phi_i=\phi_i(r),\quad\quad\quad A=A_{It}(r)\,dt\,.
\end{split}
\end{eqnarray}
Substituting the ansatz into the equations of motion~\eqref{eomphi},~\eqref{eomA} and~\eqref{eomg}, one obtains the concrete equations of motion for each field.

\begin{equation}\label{eqphi}
\frac{1}{\sqrt{BD}C}\left(\sqrt{\frac{D}{B}}Ce^{\vec\tau_i\cdot\vec\phi}\phi_i'\right)'-\frac{1}{2B}\sum_{j=1}^M e^{\vec\tau_j\cdot\vec\phi}\tau_{ji} \phi_j'^2+\frac{1}{2BD}\sum_{I=1}^N e^{\vec\gamma_I\cdot\vec\phi}\gamma_{Ii} A_{It}'^2-V_0\delta_ie^{-\vec\delta\cdot\vec\phi}=0\,,
\end{equation}
\begin{equation}\label{eqA}
\left(e^{\vec\gamma_I\cdot\vec\phi}\frac{C}{\sqrt{BD}}A_{It}'\right)'=0\,,
\end{equation}
\begin{equation}\label{eqmetric3}
\frac{2D''}{D}-\frac{2C''}{C}-\left(\frac{B'}{B}-\frac{C'}{C}+\frac{D'}{D}\right)\frac{D'}{D}+\frac{B'C'}{BC}-\frac{2}{D}\sum_{I=1}^N e^{\vec\gamma_I\cdot\vec\phi}A_{It}'^2=0\,,
\end{equation}
\begin{equation}\label{eqmetric2}
\frac{2C''}{C}-\left(\frac{B'}{B}+\frac{C'}{C}+\frac{D'}{D}\right)\frac{C'}{C}+\sum_{i=1}^M e^{\vec\tau_i\cdot\vec\phi}\phi_i'^2=0\,,
\end{equation}
\begin{equation}\label{eqmetric1}
\frac{D'C'}{DC}+\frac{1}{2}\frac{C'^2}{C^2}-{1\over 2}\sum_{i=1}^M e^{\vec\tau_i\cdot\vec\phi}\phi_i'^2+\frac{1}{2D}\sum_{I=1}^N e^{\vec\gamma_I\cdot\vec\phi}A_{It}'^2-B V_0 e^{-\vec\delta\cdot\vec\phi}=0\,.
\end{equation}
Here we have used primes to denote radial derivatives. $\tau_{ji}$ and $\gamma_{Ii}$ denote the $i$-th component of the vectors $\vec\tau_j$ and $\vec\gamma_I$, respectively.

\section{General Scaling Solutions}\label{solutions}
We are interested in the hyperscaling violation geometry with the following scaling ansatz
\begin{equation}\label{hsvscalars}
ds^2=r^{\theta}\left[-{dt^2\over r^{2z}}+{L^2\,dr^2+d\vec x^2\over r^2}\right],\quad \vec\phi=\vec \kappa\,\log r,\quad A_{It}= A_{It}(r)\,,
\end{equation}
where $\vec\kappa$ is a constant $M$-vector.

Substituting the above ansatz into~\eqref{eqA}, we find
\begin{equation}
A_{It}''-\frac{1-z-\vec\gamma\cdot\vec\kappa}{r}A_{It}'=0,\quad I=1,\dots,N
\end{equation}
from which we can determine $A_{It}$:
\begin{equation}\label{soluA}
A_{tI}(r)=\mu_I+Q_I~ r^{2-z-\vec \g_I\cdot \vec \kappa},\quad \vec \gamma_I\cdot \vec \kappa\not\neq 2-z\,.
\end{equation}
Here $\mu_I$, $Q_I$ are integration constants. If one of the charge is zero, say $Q_J=0$, then the corresponding vector field $A_J$ is trivial and the resulted theory is described by the same theory as~\eqref{action} but with $(N-1)$ vectors. It gives nothing new and therefore we consider the case with all charges non-vanishing.
There is a special case with $\vec\gamma_I\cdot \vec\kappa=2-z$ in which we instead obtain a gauge field that is logarithmically running.  If we consider the special case $\vec\gamma_I\cdot \vec\kappa=2-z$ for which $A_{tI}(r)=\mu_I+Q_I \log r$ and assume this kind of solutions dominate the geometry, from~\eqref{eqmetric3} we can easily find that $Q_I=0$.

From~\eqref{eqmetric3} we obtain
\begin{equation}\label{const2}
2(z-1)(z+2-\theta)-\sum_{I=1}^N r^{4-\theta-\vec\gamma_I\cdot\vec \kappa}(z-2+\gamma_I\cdot\vec \kappa)^2Q_I^2=0\,.
\end{equation}
Since we want to keep all charges $Q_I$ non-vanishing and in order for them to contribute at the same order, the natural way to satisfy~\eqref{const2} is to set
\begin{equation}\label{constgam}
\vec\gamma_I\cdot\vec \kappa=4-\theta\,,\quad\quad \forall ~I\,,
\end{equation}
from which we further obtain
\begin{equation}\label{const2a}
\sum_{I=1}^N Q_{I}^2=\frac{2(z-1)}{z+2-\theta}\,.
\end{equation}
Notice that the condition $\vec \gamma_I\cdot \vec \kappa\not\neq 2-z$ demands $(\theta-z-2)\neq 0$.

From~\eqref{eqmetric2} we obtain
\begin{equation}\label{const1}
(\theta-2)(\theta+2-2z)-\sum_{i=1}^M r^{\vec\tau_i\cdot\vec \kappa}\kappa_i^2=0\,,
\end{equation}
where $\kappa_i$ is the $i$-th component of the vector $\vec\kappa$. We use $A,B,\dots$ to denote the position of non-zero components of $\vec\kappa$ and $a,b,\dots$ the position of zero components of $\vec\kappa$. Then depending on whether the component $\kappa_i$ is zero or not, we distinguish two cases:
\begin{itemize}
  \item In the position $A$ of $\vec\kappa$ with the component $\kappa_{A}\neq 0$, we should demand
 \begin{equation}\label{constA}
\vec\tau_A\cdot\vec \kappa=0\,.
\end{equation}
  \item In the position $a$ of $\vec\kappa$ with $\kappa_a=0$,  there is no constraint on $\vec\tau_a\cdot\vec \kappa$ from~\eqref{const1}.
\end{itemize}
Then we have the relation
\begin{equation}\label{const1a}
\sum_{A} \kappa_A^2=\sum_{i=1}^M \kappa_i^2=\vec\kappa^2=(\theta-2)(\theta+2-2z)\,.
\end{equation}

Taking advantage of~\eqref{const2} and~\eqref{const1}, we can find the following relation by using~\eqref{eqmetric1}.
\begin{equation}\label{const3a}
L^2 \,V_0\, r^{\theta-\vec\delta\cdot\vec\kappa}-(z-\theta+1)(z-\theta+2)=0\,,
\end{equation}
which then gives
\begin{equation}\label{const3}
\vec\delta\cdot\vec\kappa=\theta,\quad L^2=\frac {(z-\theta+1)(z-\theta+2)}{V_0}\,.
\end{equation}

Using~\eqref{soluA},~\eqref{constgam} and~\eqref{const3}, the scalar equations~\eqref{eqphi} give
\begin{equation}\label{constphi}
2(\theta-z-2+\vec\tau_i\cdot\vec\kappa)r^{\vec\tau_i\cdot\vec\kappa}\,\kappa_i -\sum_{j=1}^M r^{\vec\tau_j\cdot\vec\kappa}\tau_{ji}\,\kappa_j^2 -2L^2\,V_0\, \delta_i+(\theta-z-2)^2\sum_{I=1}^N Q_I^2\, \gamma_{Ii}=0\,.
\end{equation}
They are $M$ coupled equations and can be further simplified in the following way:
\begin{itemize}
  \item If we choose $i$ such that $\kappa_i\neq 0$, we obtain
\begin{equation}\label{constphi1}
\begin{split}
& \vec\tau_A\cdot\vec \kappa=\sum_A \tau_A \kappa_A=0\,,\\
\kappa_A-(\theta-z-1)\, \delta_A&+\frac{(\theta-z-2)}{2}\sum_{I=1}^N Q_I^2\, \gamma_{IA}- \frac{1}{2(\theta-z-2)}\sum_{B} \tau_{BA}\,\kappa_B^2=0\,.
\end{split}
\end{equation}
Note that we have used $A,B$ to denote the position of the non-zero components of $\vec\kappa$. If there is only one scalar in which the vector $\vec\tau_1$ is now a constant number $\tau_1$, the way to satisfy the above relations is to set $\tau_1=0$.

  \item On the other hand, when $i=a$ with $\kappa_a=0$, we obtain
 \begin{equation}\label{constphi2}
-(\theta-z-1)\, \delta_a+\frac{(\theta-z-2)}{2}\sum_{I=1}^N Q_I^2\, \gamma_{Ia}- \frac{1}{2(\theta-z-2)}\sum_{B} \tau_{Ba}\,\kappa_B^2=0\,.
\end{equation}
\end{itemize}
Two sets of relations~\eqref{constphi1} and~\eqref{constphi2} are non-linear equations for $\kappa_i$ after we introduce non-trivial metric $\mathcal{G}_{ii}(\phi)$ into the action~\eqref{action}.

To sum up, we obtain the hyperscaling violating geometry if there is a consistent solution of $(z,\theta, L^2, Q_I, \vec\kappa)$ that satisfy
\begin{equation}\label{allconstaints}
\begin{split}
\vec\gamma_I\cdot\vec \kappa=4-\theta\,, \forall ~I\,,\quad \vec\delta\cdot\vec\kappa=\theta\,,\quad \sum_{I=1}^N Q_{I}^2=\frac{2(z-1)}{z+2-\theta}\,,\\
\vec\kappa^2=(\theta-2)(\theta+2-2z)\,,\quad L^2=\frac {(z-\theta+1)(z-\theta+2)}{V_0}\,,
\end{split}
\end{equation}
as well as constraints from scalar equations~\eqref{constphi1} and~\eqref{constphi2}. Note that in the presence of $M$ vectors $\vec\tau_i$ we in general do not have an analytic solution.

\subsection{Scaling solutions with $\vec\tau_1=...=\vec\tau_M$}
In this part we focus on a simple case with $\vec\tau_1=...=\vec\tau_M=\vec\tau$. Then from~\eqref{const1} we obtain
 \begin{equation}\label{constau}
\vec\tau\cdot\vec \kappa=0\,.
\end{equation}
Using this condition as well as~\eqref{allconstaints}, the scalar equations~\eqref{constphi} are deuced to
\begin{equation}\label{conskappa}
\vec \kappa =(\theta-z-1)\vec\delta-{1\over 2}(\theta-z-2)\sum_I Q_I^2~\vec \gamma_I+\frac{(\theta-2)(\theta+2-2z)}{2(\theta-z-2)}\vec\tau\,.
\end{equation}
Substituting $\vec\kappa$ into the first two equations of~\eqref{allconstaints} and using~\eqref{constau}, we obtain the following relations.
\begin{equation}\label{eq1}
(\theta-z-1)\vec\delta\cdot\vec\delta-{1\over 2}(\theta-z-2)\sum_I Q_I^2~\vec \gamma_I \cdot  \vec\delta+\frac{(\theta-2)(\theta+2-2z)}{2(\theta-z-2)}\vec\tau\cdot\vec\delta=\theta\,,
\end{equation}
\begin{equation}\label{eq2}
(\theta-z-1)\vec\delta\cdot\vec\gamma_I-{1\over 2}(\theta-z-2)\sum_J Q_J^2~\vec \gamma_I \cdot  \vec\gamma_J+\frac{(\theta-2)(\theta+2-2z)}{2(\theta-z-2)}\vec\tau\cdot\vec\gamma_I=4-\theta\,,
\end{equation}
\begin{equation}\label{eqtau}
(\theta-z-1)\vec\delta\cdot\vec\tau-{1\over 2}(\theta-z-2)\sum_I Q_I^2~\vec \gamma_I \cdot  \vec\tau+\frac{(\theta-2)(\theta+2-2z)}{2(\theta-z-2)}\vec\tau\cdot\vec\tau=0\,,
\end{equation}
where the last one is from~\eqref{constau} and should be considered as a non-trivial constraint on $\vec\tau$. We can in principle determine $(z,\theta, Q_I)$ from above three sets of conditions together with
\begin{equation}\label{const2b}
\sum_{I=1}^N Q_{I}^2=\frac{2(z-1)}{z+2-\theta}\,.
\end{equation}

To solve those equations we introduce
\begin{equation}
X_{IJ}=\vec\gamma_I \cdot  \vec\gamma_J,\quad X_I=\vec\delta\cdot\vec\gamma_I,\quad Y_I=\vec\tau\cdot\vec\gamma_I,\quad X_0=\vec\delta\cdot\vec\delta\,,\quad Y_0=\vec\tau\cdot\vec\delta\,,\quad Z_0=\vec\tau\cdot\vec\tau\,,
\end{equation}
and assume that the matrix $X_{IJ}$ is reversible with its inverse $X^-_{IJ}$. Then the charge $Q_I$ can be determined from~\eqref{eq2}:
\begin{equation}\label{eqQ}
Q_I^2=\frac{2(\theta-z-1)}{\theta-z-2}\sum_J X^-_{IJ}X_J+\frac{(\theta-2)(\theta-2z+2)}{(\theta-z-2)^2}\sum_J X^-_{IJ}Y_J+\frac{2(\theta-4)}{\theta-z-2}\sum_J X^-_{IJ}\,.
\end{equation}

On the other hand, in the case where $X_{IJ}$ does not have an inverse, for each zero eigenvector $\vec\xi$, we obtain an equation on $(\theta,z)$ which does not involve the charges:
\begin{equation}\label{zeroeigen}
(z+1-\theta)\sum_{I=1}^N X_I\xi_I=\frac{(\theta-2)(\theta+2-2 z)}{2}\sum_{I=1}^N Y_I\xi_I+(\theta-4)\sum_{I=1}^N \xi_I,
\end{equation}
with $\xi_I$ the $I$-th component of $\vec\xi$. Nothing special happens in the case of a single zero eigenvector and~\eqref{zeroeigen} is one of the constraints on $(\theta,z)$. If the matrix $X_{IJ}$ has two or more eigenvectors, unless the ratios ${\sum_I X_I\xi_I\over \sum_I\xi_I}, {\sum_I Y_I\xi_I\over \sum_I\xi_I}$
are the same for all such zero eigenvectors, the only possible way is to choose $(\theta=4, z=3)$. However, such value should be excluded because it gives $\vec\kappa^2=0$ as can be seen from~\eqref{allconstaints}.

Substituting~\eqref{eqQ} into~\eqref{eq2},~\eqref{eqtau} and~\eqref{const2b}, we obtain
\begin{equation}\label{eq3}
(\theta-z-1) W_1+\frac{(\theta-2)(\theta-2z+2)}{2(\theta-z-2)}V_1+(\theta-4)W_0=1-z\,,
\end{equation}
\begin{equation}\label{eq4}
(4-\theta)W_1-(\theta-z-1)(W_2-X_0)-\frac{(\theta-2)(\theta-2z+2)}{2(\theta-z-2)}(V_2-Y_0)=\theta\,,
\end{equation}
\begin{equation}\label{eqtaua}
(4-\theta)V_1-(\theta-z-1)(V_2-Y_0)-\frac{(\theta-2)(\theta-2z+2)}{2(\theta-z-2)}(U_2-Z_0)=0\,.
\end{equation}
Here we have defined the following scalar combination:
\begin{equation}
\begin{split}
W_0=\sum_{I,J}X^-_{IJ},\quad W_1=\sum_{I,J}X^-_{IJ} X_J,\quad W_2=\sum_{I,J}X_I X^-_{IJ} X_J\,,\\
 V_1=\sum_{I,J}X^-_{IJ} Y_J ,\quad V_2=\sum_{I,J}X_I X^-_{IJ} Y_J\,,\quad U_2=\sum_{I,J}Y_I X^-_{IJ} Y_J\,.
\end{split}
\end{equation}

Two scaling components $(\theta,z)$ can be solved exactly from~\eqref{eq3} and~\eqref{eq4}. However, the solutions are too complicated to give any useful information.  We stress out that the resulted solution should further satisfy~\eqref{constau} or equivalently~\eqref{eqtaua}.  This in turn imposes some constraint on the choice of $\vec\tau$.
It is difficult to determine the exact conditions for $\vec\tau$ such that the value of $(\theta,z)$ solved from~\eqref{eq3} and~\eqref{eq4} is real and other equations are compatible with this value. We instead give a simple choice of $\vec\tau$ as an example in the next section.
\section{Analytic Solutions}\label{analytic}
After establishing general results on the properties of purely hyperscaling violating solutions, we apply the previous formulae to the special case by choosing $\vec\tau=0$. We will present a set of exact hyperscaling violating solutions and then discuss in details the physical constraints on $(\theta,z)$ in order to give a well defined holographic ground state solution.

Two equations~\eqref{eq3} and~\eqref{eq4} are simplified significantly as $\vec\tau=0$, which can be written as

\begin{equation}\label{matrix}
\begin{pmatrix}
W_1+W_0 & 1-W_1\\
W_2+W_1-X_0+1 & X_0-W_2
\end{pmatrix}
\begin{pmatrix}
\theta\\
z
\end{pmatrix}=
\begin{pmatrix}
      W_1+4W_0+1   \\
      W_2+4W_1-X_0
\end{pmatrix}\,.
\end{equation}
We find that the non-linear terms of $(\theta,z)$ disappear and the constraint~\eqref{eqtaua} becomes trivial as $\vec\tau=0$.
The solution is given by
 \begin{equation}\label{solu}
 \begin{split}
 \theta=&2\frac{2W_1(1-W_1)+(1+2W_0)W_2-X_0(1+2W_0)}{1-W_1^2+(1+W_0)W_2-X_0(1+W_0)}\,, \\
z=&\frac{1+4W_0+2W_1+W_2-3W_1^2+3W_0W_2-X_0(1+3W_0)}{1-W_1^2+(1+W_0)W_2-X_0(1+W_0)}\,,
\end{split}
\end{equation}
when the coefficient matrix in the left hand side of~\eqref{matrix} is reversible, which means the denominator in~\eqref{solu} is non-vanishing. Otherwise, there is no solution of~\eqref{matrix}.

After obtaining $(\theta,z)$ as given in~\eqref{solu}, other parameters of the hyperscaling violating solution~\eqref{hsvscalars}, $(\vec\kappa, L^2, Q_I)$, can be determined from~\eqref{allconstaints} and
\begin{equation}
\begin{split}
\vec \kappa &=(\theta-z-1)\vec\delta-{1\over 2}(\theta-z-2)\sum_I Q_I^2~\vec \gamma_I\,,\\
Q_I^2&=\frac{2(\theta-z-1)}{\theta-z-2}\sum_J X^-_{IJ}X_J+\frac{2(\theta-4)}{\theta-z-2}\sum_J X^-_{IJ}\,.
\end{split}
\end{equation}
 So far, a kind of exact hyperscaling violating solutions are in principle completely determined.

\subsection{Constraints on the parameter space of $(\theta,z)$}
Note that the hyperscaling violating geometry~\eqref{hsvscalars} suffers from curvature and null singularities for general values of $(\theta,z)$. To ensure that the background can be taken to describe a well defined ground state, we should impose some necessary constraints to restrict the parameter space of the scaling exponents (or equivalently theory parameters).

\begin{figure}[ht!]
\begin{center}
\includegraphics[width=.5\textwidth]{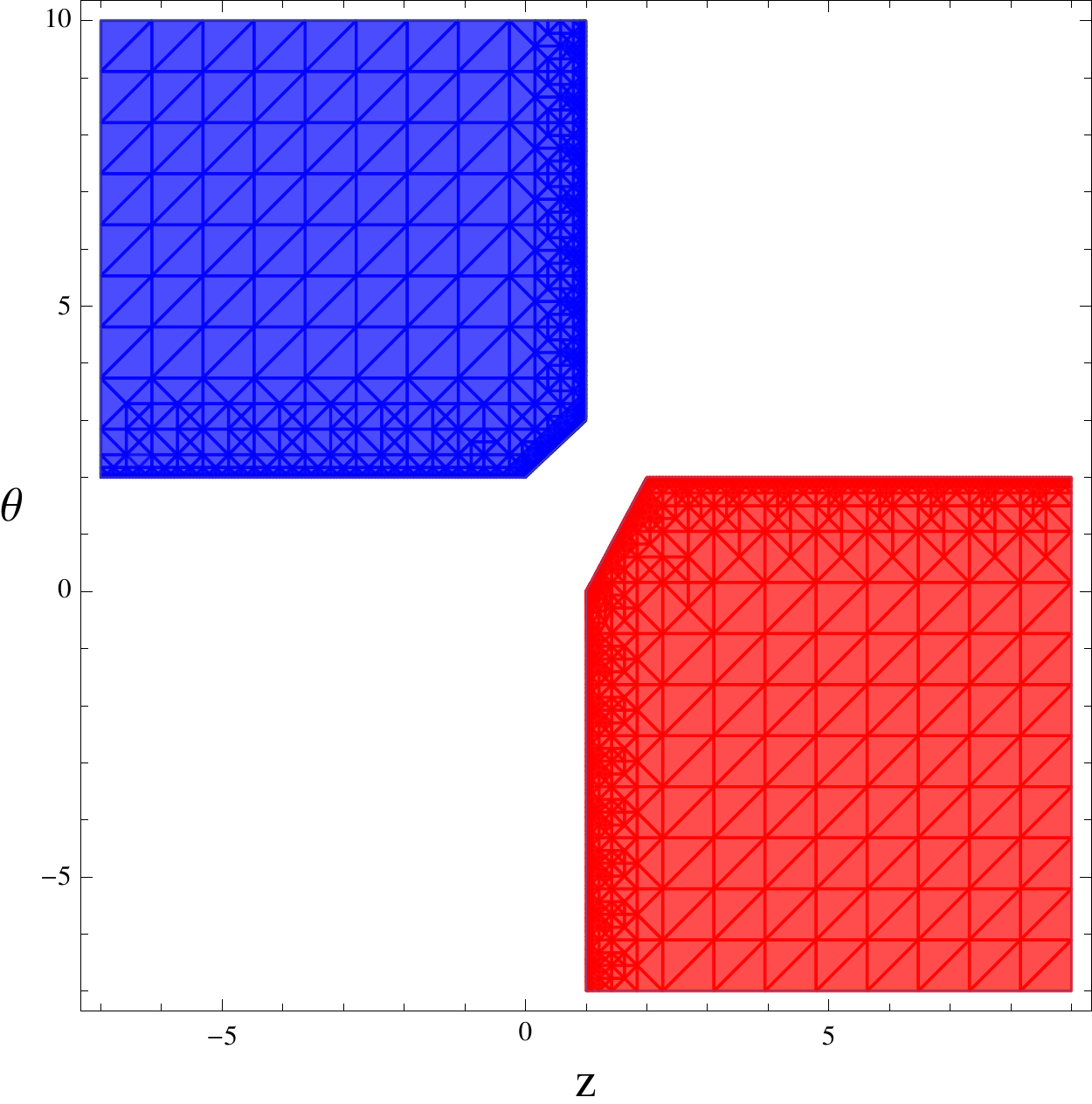}
\caption{Parameter space for $\theta$ and $z$ that satisfies all the constraints to have a well defined holographic ground state solution. The blue region on the upper left corresponds to the geometry with the IR located at $r\rightarrow 0$, and the lower right red region is the case with IR at $r\rightarrow \infty$. }
\label{fig:range}
\end{center}
\end{figure}

\begin{itemize}
  \item To have a well defined solution, $(L^2, \vec\tau^2, Q_I^2)$ of~\eqref{allconstaints} should be positive, which demands
\begin{equation}\label{positive}
(z-\theta+1)(z-\theta+2)> 0,\quad (z-1)(z+2-\theta)>0,\quad (\theta-2)(\theta+2-2z)>0\,.
\end{equation}
It is reasonable to impose the null energy condition. However, it turns out that the condition~\eqref{positive} is much stronger than the null energy condition.\,\footnote{The null energy condition for the hyperscaling violation geometry~\eqref{hsvscalars} is given by
\begin{equation}
(z-1)(z+2-\theta)\geqslant0,\quad (\theta-2)(\theta+2-2z)\geqslant 0\,.
\end{equation}
}

  \item We require the $(t,\vec x)$ components of the metric to scale the same way with $r$
\begin{equation}\label{irgood}
(\theta-2)(\theta-2z)>0\,,
\end{equation}
such that the IR region is unambiguous. The IR can be located at either $r\rightarrow 0$  or $r\rightarrow\infty$, which depends on where the $(t,\vec x)$ metric elements vanish:
\begin{equation}\label{IRgood}
\begin{split}
\text{IR}\;\;\quad r\rightarrow 0:&\; \theta>2,\quad \theta>2z\,,\\
\text{IR}\quad r\rightarrow \infty:&\; \theta<2,\quad \theta<2z\,.
\end{split}
\end{equation}

  \item In order to resolve the singularity, we follow Gubser's physicality criterion~\cite{Charmousis:2010zz,Gubser:2000nd}, which requires the temperature deformation of the metric~\eqref{hsvscalars} to be relevant. The temperature deformation in present case with $\vec\tau=0$ can be obtained as follows. We turn on the temperature deformation as
\begin{equation}\label{finitT}
ds^2=r^\theta\left(-f(r)\frac{dt^2}{r^{2z}}+\frac{L^2 dr^2}{r^2 f(r)}+\frac{dx^2+dy^2}{r^2}\right),
\end{equation}
and find $f(r)$. By substituting~\eqref{finitT} into the equations of motion~\eqref{eqphi}-\eqref{eqmetric1} and using the relations~\eqref{allconstaints} as well as $\vec\tau=0$, we obtain the equation of motion for $f(r)$,
\begin{equation}\label{eomf}
f'(r)-\frac{z+2-\theta}{r}f(r)+\frac{z+2-\theta}{r}=0\,,
\end{equation}
from which we can determine its configuration
\begin{equation}\label{temdeform}
f(r)=1+\zeta\;r^{2+z-\theta}\,,
\end{equation}
with $\zeta$ an arbitrary constant. Demanding the temperature deformation~\eqref{temdeform} to be relevant, we arrival at  the following constraint:
\begin{equation}\label{Tdeform}
\begin{split}
\text{IR}\;\;\quad r\rightarrow 0:&\; 2+z-\theta<0\,,\\
\text{IR}\quad r\rightarrow \infty:&\; 2+z-\theta>0\,.
\end{split}
\end{equation}
\end{itemize}
The allowed parameter space is given by combing all above conditions:
\begin{equation}\label{range}
\begin{split}
\text{IR}\quad r\rightarrow 0&: [z\leqslant 0, \theta>2],\quad [0<z< 1, \theta>z+2]\,,\\
\text{IR}\quad r\rightarrow \infty&: [1< z\leqslant 2, \theta< 2z-2],\quad [z> 2, \theta<2]\,.
\end{split}
\end{equation}
We present the corresponding parameter range in figure~\ref{fig:range}. 

For a black brane solution with the horizon located at $r_h$, one should demand $f(r_h)=0$. Therefore, we can fix the integration constant,
\begin{equation}\label{heat}
f(r)=1-\left(\frac{r}{r_h}\right)^{2+z-\theta}.
\end{equation}
The temperature associated with~\eqref{heat} is given by
\begin{equation}
T=\frac{|z+2-\theta|}{4\pi L}r_h^{-z}.
\end{equation}
We stress out that the black brane solution~\eqref{finitT} with~\eqref{heat} is valid only for the case $\vec\tau=0$. In others cases with general $\vec\tau_i$, although the black brane solution~\eqref{finitT},~\eqref{heat} satisfies the Einstein equations~\eqref{eqmetric3}-\eqref{eqmetric1} and vector equations~\eqref{eqA}, it breaks down in the scalar equations~\eqref{eqphi} due to the non-trivial scalar couplings with the kinetic term for each scalar $\phi_i$. 

In the deep IR each vector $A_I$  generates a non-zero flux,
\begin{equation}
\frac{1}{4\pi}\int_{R^2}Z_I(\phi)\;{}^\star F_I=-\frac{\omega_{(2)}}{4\pi} Z_I(\phi)\frac{C(r) A_{It}'}{\sqrt{B(r)D(r)}}=\frac{z+2-\theta}{4\pi L}Q_I\, \omega_{(2)}\,,
\end{equation}
with $\omega_{(2)}$ the volume of the spatial section of the metric. Therefore, the hyperscaling violating geometry we obtained with $Q_I\neq 0$ describes the quantum criticality at fractionalized phases.

\subsection{A top-down example}
\label{subsec:top}
As a simple example, in this part we consider a top-down setup. We begin with a D-dimensional closed string action ($D\geqslant 5$) in the string frame~\cite{elias},
\begin{equation}
S_{closed}=\int d^Dx\sqrt{g_{\s}}~e^{-2\Phi}~\Big[{\cal R}+4(\nabla \Phi)^2-\frac{H^2}{12}-\frac{e^{2\Phi}}{4}F_c^2+V_0+\cdots\Big]\,,
\end{equation}
where $H=dB_2$ is the field strength of the NS two-form and $F_c=dC$ is a RR field strength. 

On compactifing the theory to 4 dimensions on a manifold with linear dimension $R$, we obtain
\begin{equation}\label{compact}
S_{closed}=\int d^4x\sqrt{g_{\s}}~e^{-\chi}~\Big[{\cal R}+(\nabla \chi)^2-\frac{(\nabla R)^2}{R^2}-\frac{R^2}{4}F_g^2-{1\over 4}F_B^2-\frac{e^{\chi}R^{D-4}}{4}F_c^2+V_0+\cdots\Big]\,,
\end{equation}
where we have introduced
\begin{equation}
 e^{-\chi}=R^{D-4}\;e^{-2\Phi}\,.
\end{equation}
Mapping~\eqref{compact} to the Einstein frame
\begin{equation}
g_{\s}=e^{\chi}\;g_{E}\,,
\end{equation}
and defining
\begin{equation}
\phi_1=\frac{1}{\sqrt{3}}(-\chi+2\log{R}),\quad \phi_2=\sqrt{\frac{2}{3}}(\chi+\log R)\,,
\end{equation}
we arrival at the theory
\begin{equation}\label{string}
\begin{split}
S_{closed}=\int d^4x\sqrt{g_{E}}\Big[{\cal R}_E-\frac{1}{2}\left[(\pa\phi_1)^2+(\pa\phi_2)^2\right]-{e^{\sqrt{3}\phi_1}\over 4}F_g^2-{e^{\frac{\phi_1-\sqrt{2}\phi_2}{\sqrt{3}}}\over 4}F_B^2\\
-{e^{\frac{D-4}{\sqrt{6}}(\sqrt{2}\phi_1+\phi_2)}\over 4}F_c^2+V_0~e^{\frac{-\phi_1+\sqrt{2}\phi_2}{\sqrt{3}}}+\cdots\Big]\,.
\end{split}
\end{equation}

So we obtain a top-down theory with two scalars and three massless vectors. We can apply our previous discussion to look for hyperscaling violating geometries in present setup~\eqref{string}. We find that the scaling solution with three non-vanishing charges associated with three vectors is not allowed. However, we do obtain non-trivial hyperscaling violating geometries by consistently setting $F_g=F_B=0$. More precisely, we obtain
\begin{equation}\label{topvalue}
\theta=D^2-8 D+20,\quad z=-1\,,
\end{equation}
and
\begin{equation}
\begin{split}
\vec\kappa=\left(\frac{3+D(z-1)-5 z+\theta}{\sqrt{3}},\frac{D(z-1)-2(\theta+z-3)}{\sqrt{6}}\right)\,,\\
L^2=\frac{(D^2-8 D+19)(D^2-8 D+20)}{V_0},\quad Q_c^2=\frac{4}{D^2-8D+19}\,.
\end{split}
\end{equation}
We notice that the values of $(\theta,z)$ given in~\eqref{topvalue} satisfy the Gubser bound as shown in figure~\ref{fig:range}. 

\section{Conclusion and Discussion}\label{conclusion}
In this paper we have considered a generalised Einstein-Maxwell-Dilaton system which involves gravity coupled with an arbitrary number of scalars and vector fields. We discussed the hyperscaling violating geometry in this theory and derived the general conditions for the existence of such quantum critical geometry driven by running scalars.

In a particular case where the kinetic term of each scalar takes the standard form, $\frac{1}{2}(\partial\phi)^2$,  we can obtain a set of analytic solutions with arbitrary hyperscaling violation exponent $\theta$ and dynamical exponent $z$, see~\eqref{solu}. The corresponding black brane solutions can be found exactly by turning on a blackness function~\eqref{heat}.
Since the extreme geometry~\eqref{hsvscalars} has typically naked singularities, we also incorporated the development of the Gubser criterion for the acceptability of a naked singularity.
The allowed parameter range for $(\theta,z)$ to resolve the singularity is presented in~\eqref{range} and figure~\ref{fig:range}. As one can see that in the solutions we obtained  the charges are fully determined by $(\vec \gamma_I,\vec \delta, V_0)$  and therefore are not tunable. These solutions could at best be thought of as the near extremal approximations to general solutions.~\footnote{Apart from the quantum critical geometry driven by running scalars, there is also a special geometry, i.e., $AdS_2\times \mathbb{R}^2$ with constant scalars, which corresponds to $z\rightarrow \infty$ with $\theta$ finite. Such IR geometry in a theory with one scalar and several vectors was discussed in~\cite{Tarrio:2011de,Alishahiha:2012qu}.}

We have focused on the hyperscaling violating geometries and therefore considered the scalar potential and kinetic functions that are chosen to be simple exponentials. Nevertheless, to adopt the standard holographic dictionary and ensure a UV CFT, one should embed these solutions in $AdS$ spacetime. Therefore, the full background is described by gravitational solutions that are hyperscaling violation in the IR and asymptotically approach  AdS in the UV. This can be done by modifying the coupling functions appropriately such that the scalars settle down to a constant at the $AdS$ boundary.

In this paper we considered an effective theory with many scalars and vectors at the two-derivative level. We limit ourselves to a simple scalar manifold, which limits its applicability in the supergravity context. It is known that in supergravity context scalars often live on manifolds endowed with some group-theoretic structure. We gave a simple top-down example in subsection~\ref{subsec:top}. However, in order to allow a robust set of embeddings into top-down models, it is interesting to consider the most general scalar manifold. For instance, it is of some interest to choose one that appears in supergravity to look for analytic solutions that dual to hyperscaling phases in ABJM theory~\cite{DeWolfe:2014ifa}.~\footnote{We thank the referee for pointing out this point.} We will leave them for future studies.


\addcontentsline{toc}{section}{Acknowledgments}
\acknowledgments\label{ACKNOWL}

LL would like to thank E.~Kiritsis for useful discussions and comments.

This work was supported in part by European Union's Seventh Framework Programme under grant agreements (FP7-REGPOT-2012-2013-1) no 316165.




\end{document}